# Spin-charge conversion in NiMnSb Heusler alloy films

Zhenchao Wen,[1,2][*][†] Zhiyong Qiu,[3,1][†] Sebastian Tölle,[4][‡] Cosimo Gorini,[5][§] Takeshi Seki,[1,2] Dazhi Hou,[6] Takahide Kubota,[1,2] Ulrich Eckern,[4] Eiji Saitoh,[1,6] Koki Takanashi[1,2]

[1]*Institute for Materials Research, Tohoku University, Sendai 980-8577, Japan*
[2]*Center for Spintronics Research Network, Tohoku University, Sendai 980-8577, Japan*
[3]*Key Laboratory of Materials Modification by Laser, Ion, and Electron Beams (Ministry of Education), School of Materials Science and Engineering, Dalian University of Technology, Dalian, China*
[4]*Universität Augsburg, Institut für Physik, 86135 Augsburg, Germany*
[5]*Universität Regensburg, Institut für Theoretische Physik, 93040 Regensburg, Germany*
[6]*WPI Advanced Institute for Materials Research, Tohoku University, Sendai 980-8577, Japan*

**Abstract**

Half-metallic Heusler alloys are attracting considerable attention because of their unique half-metallic band structures which exhibit high spin polarization and yield huge magnetoresistance ratios. Besides serving as ferromagnetic electrodes, Heusler alloys also have the potential to host spin-charge conversion which has been recently demonstrated in other ferromagnetic metals. Here, we report on the spin-charge conversion effect in the prototypical Heusler alloy NiMnSb. Spin currents were injected from $Y_3Fe_5O_{12}$ into NiMnSb films by spin pumping, and then the spin currents were converted to charge currents via spin-orbit interactions. Interestingly, an unusual charge signal was observed with a sign change at low temperature, which can be manipulated by film thickness and ordering structure. It is found that the spin-charge conversion has two contributions. First, the interfacial contribution causes a negative voltage signal, which is almost constant versus temperature. The second contribution is temperature dependent because it is dominated by minority states due to thermally excited magnons in the bulk part of the film. This work provides a pathway for the manipulation of spin-charge conversion in ferromagnetic metals by interface-bulk engineering for spintronic devices.

[*]wen.zhenchao@imr.tohoku.ac.jp
[‡]sebastian.toelle@physik.uni-augsburg.de, [§]cosimo.gorini@physik.uni-regensburg.de
[†] These authors contributed equally to this work.

**INTRODUCTION**

The spin-charge conversion in well-designed materials and/or new states of matter is essential for the development of future energy-efficient spintronic devices (*1–7*). Recently, spin currents generated from ferromagnets (FMs), such as CoFeB (*8, 9*), NiFe (*9–11*), and FePt (*12*), is attracting great attention not only due to the remarkable spin signals but also due to their controllability owing to interactions between spin and magnetization. Among FMs, half-metallic ferromagnets (HMFs) are a unique class of FMs with respect to their electronic band structure: one spin channel exhibits a band gap at the Fermi level, and the other one is conductive (*13, 14*). This unique band structure yields a high spin polarization (ideally 100%), resulting in very high performances of HMF-based spintronic devices, such as a huge tunneling magnetoresistance (TMR) ratio of more than 2000% (*15*). However, the spin-charge conversion, a basic spin-related effect, was considered to be forbidden because of only one spin channel is available in the ground state of HMFs. Ohnuma *et al.* theoretically predicted that spin-charge conversion can occur with the assistance of magnons at a finite temperature (*16*), which opens a new possibility for spin-charge conversion in HMFs. Furthermore, the degradation of perfect half-metallicity in the interface/surface region (*17–22*) is another possibility of spin-charge conversion in these materials.

Here, we study the spin-charge conversion in a NiMnSb Heusler alloy which was first reported to be a HMF by first-principles calculations (*13*). Figure 1**A** shows a schematic illustration of its crystal structure in the $C1_b$ phase. Recently, Wen *et al.* succeeded in measuring a current-perpendicular-to-plane (CPP) giant magnetoresistance (GMR) with high-quality NiMnSb films (*23, 24*). Ciccarelli *et al.* reported on spin-orbit torques in NiMnSb at room temperature (*25*). As shown in Fig. 1**B**, NiMnSb possesses a typical half-metallic band structure where the minority spin band shows a gap at the Fermi level for the bulk state. However, the minority gap closes owing to electron-magnon interactions and interface/surface states (Fig. 1**B**) (*17*). Thus, the spin-charge conversion is expected due to the bulk and interface contributions, as shown in Fig. 1**C**.

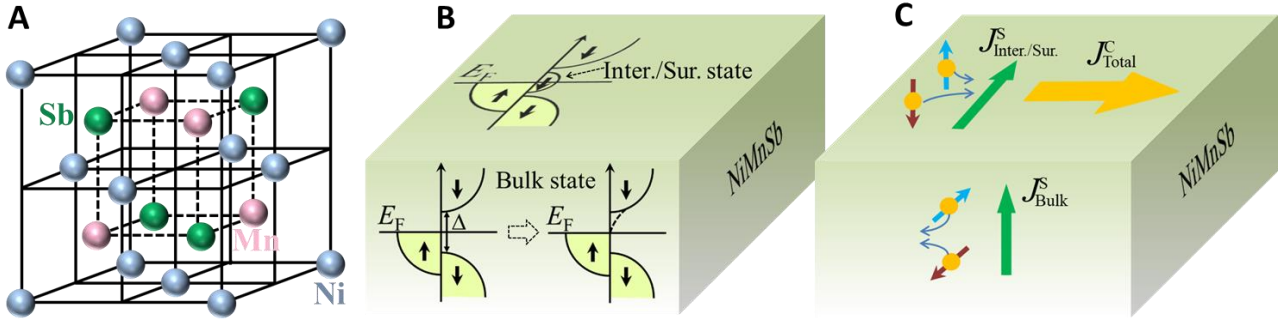

**Fig. 1. Schematic illustrations for the NiMnSb crystal and electronic band structures.** (**A**) Crystal structure of NiMnSb in the $C1_b$ phase. (**B**) Half-metallic band structure in NiMnSb. The minority band gap closes due to the interaction between electrons and magnons, i.e. righthand-side illustration of bulk state. The minority band state exists at the interface/surface (Inter./Sur.). (**C**) Spin-charge conversion with interface and bulk contributions. Here, $J^S_{Inter.Sur.}$ and $J^S_{Bulk}$ represent the spin currents due to interface/surface and bulk, respectively. $J^C_{Total}$ indicates the converted total charge current.

In this work, NiMnSb films with varying thickness and ordering structure were deposited on $Y_3Fe_5O_{12}$ (YIG) substrates, the quality of which was confirmed by structural analysis and anisotropic magnetoresistance (AMR). Spin currents were injected into NiMnSb from the insulating YIG layers by spin pumping. Then the spin currents were converted to charge currents by means of spin-orbit interactions and detected as voltage signals. Interestingly, an unusual temperature dependence of the voltage was observed with a sign change at low temperature, which can be controlled by the thickness and ordering structure of the NiMnSb films. The origin of the unusual voltage signals can be well interpreted by the spin-charge conversion due to interface and bulk contributions.

**RESULTS**

Figure 2**A** shows the structural properties of NiMnSb films characterized by out-of-plane X-ray diffraction (XRD). Two kinds of samples with different ordering structure were prepared on YIG substrates. One was annealed at a high temperature of 500 °C, and the other, named as less-ordered sample, was grown at room temperature. In addition to the diffraction peaks from YIG substrate, XRD diffraction peaks from the NiMnSb (111) and (002) superlattices were observed for the NiMnSb film annealed at 500 °C. Whereas, only a weak (111) peak is shown in the XRD pattern for the sample deposited at room temperature, which indicates the as-grown sample has a low structural order. Furthermore, the magnitude of half metallic feature was investigated by AMR measurement because the AMR effect with a negative sign was reported to be necessary for examining the electronic band structures of HMFs (*26*, *27*). Figure 2**B** and 2**C** show the AMR effect for the two kinds of samples measured at 10 K and 300 K, respectively. The dependence of AMR ratio on the in-plane angle $\phi$, where $\phi$ = 0° (90°) represents that magnetization is normal (parallel) to the measuring current, shows a negative sign ($\rho_{0°} > \rho_{90°}$) for the NiMnSb film. The AMR effect at 10 K is much larger than that measured at 300 K, which indicates the reduction of the spin polarization at the high temperature due to thermal excitation. For the less-ordered sample, the AMR effect is smaller than that in the high-ordered sample which is consistent with the degradation in half metallicity.

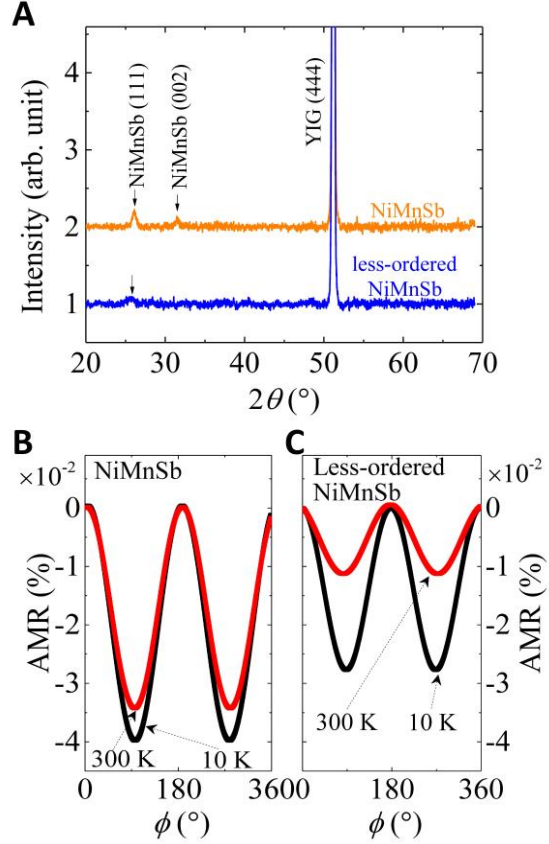

**Fig. 2. Structural properties of NiMnSb films and AMR effect in the films.** (**A**) Out-of-plane XRD patterns for 20-nm-thick NiMnSb films with two different ordering structure. (**B**) AMR effect for the two kinds of NiMnSb films measured by an in-plane $\phi$ scan method at a magnetic field of 2 T at 10 and 300 K, respectively.

Furthermore, the measurement of spin-charge conversion was carried out by a spin pumping method, as illustrated in Fig. 3**A**. Ferromagnetic resonance (FMR) is excited in YIG by applying a microwave with an external magnetic field, thus spin currents are pumped into NiMnSb. Because of the spin-orbit interactions, an electric current is generated by spin-to-charge conversion in NiMnSb, which can be picked up by the electrodes placed on the NiMnSb film. The dependence of electric voltage on temperature was investigated in the temperature range of 10−300 K. Note that the FMR absorption spectrum of NiMnSb layer and the electric voltage at the NiMnSb resonance were also observed, and the signals are much smaller than those due to YIG and can be distinct from the separation of resonance field positions of NiMnSb and YIG. The details are shown in Supplementary

Materials. Figure 3**B** indicates magnetic field dependence of voltage signal measured at 10 and 300 K under a microwave source of 5 GHz and 25 mW. Interestingly, a sign change of the voltage was observed at 10 K. Figure 3**C** shows detailed spectra for both FMR of YIG (left panel) and voltage signals in NiMnSb (right panel) measured at different temperatures. Corresponding to the resonance absorption, voltage signals were observed in the NiMnSb film grown on the YIG substrate. The magnetic field ($H_V$) at the maximum of voltage coincides with the resonance field ($H_{FMR}$) of FMR absorption, which indicates that the voltage signal originates from the FMR of YIG. In addition, since the NiMnSb is a ferromagnetic material, the anomalous Hall effect (AHE) could appear due to rectification effect (*28*), which may also contribute an electric voltage. In order to make a quantitative analysis, the observed voltage signals were fitted by the following equation (*29*):

$$V = V_{SC}\frac{\Delta H^2}{(H_{ex}-H_V)^2+\Delta H^2} + V_{AHE}\frac{-2\Delta H(H_{ex}-H_V)}{(H_{ex}-H_V)^2+\Delta H^2} \qquad (1)$$

where $V_{SC}$ represents the voltage by spin-charge conversion; $V_{AHE}$ is the contribution from AHE; $\Delta H$ is the full width at half maximum for the voltage signal. Typical fitting results for the measured voltage signals at 10 K and 300 K are shown as solid lines in Fig. 3**B**. It is found that the observed voltage is mainly contributed by spin-charge conversion, where the ratio of $V_{SC}/V_{AHE}$ is obtained to be 12.7 at 300 K (11.9 at 10 K). For the other parameters, the $H_V$ is 1020 Oe at 300 K (886 Oe at 10 K) and the $\Delta H$ is 21 Oe at 300 K (28 Oe at 10 K). Note that the parameter values were averaged from the fitting results in negative and positive magnetic field directions.

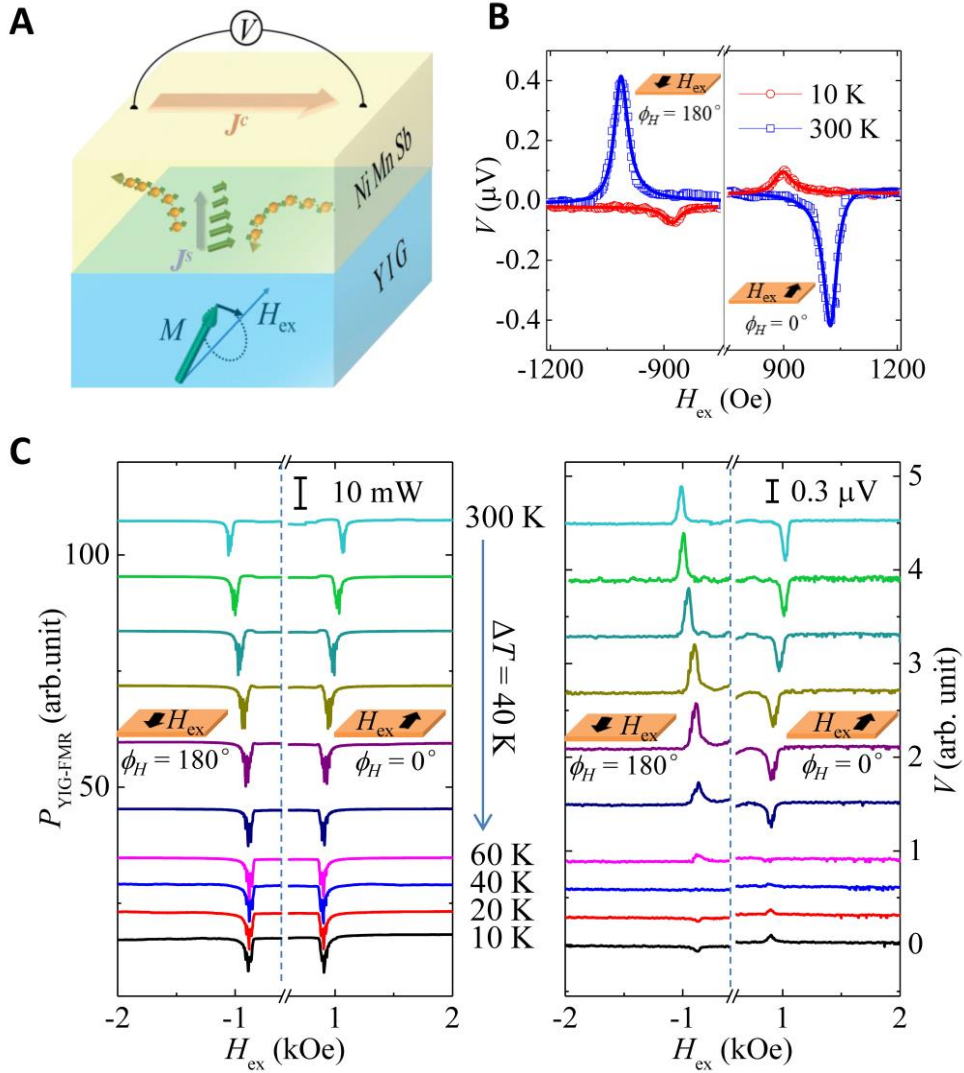

**Fig. 3. ISHE measurement with spin pumping.** (**A**) Schematic illustration of the YIG//NiMnSb sample with experiment setup carried out in the study. (**B**) Magnetic field dependence of the electronic voltage, $V$, measured at 10 and 300 K. The directions of external magnetic field, $H_{ex}$, are also indicated by the inset illustrations. The blue and red solid lines are fitting results for the experimental data by Eq. (1). (**C**) FMR spectra in YIG (left panel) and voltage signals in NiMnSb (right panel) measured in the temperature range from 10 to 300 K. The applied microwave power and frequency here are 25 mW and 5 GHz, respectively.

The dependence of $V_{SC}$ on measuring temperature was further investigated under different microwave conditions and NiMnSb thicknesses. Figure 4A shows the temperature dependence of $V_{SC}$ with varying microwave power of $P_{in}$ = 50, 100, and 200 mW at a fixed microwave frequency of $f$ = 5 GHz for a 20-nm-thick NiMnSb film. It is observed that a sign change of $V_{SC}$ appears at the temperature of $T_{cross} \sim 40$ K. With further increasing temperature, a dramatic increase of the amplitude of $V_{SC}$ is observed up to $\sim 150$ K. Then, the amplitude of the $V_{SC}$ shows a weak dependence on temperature when the measuring temperature increases to 300 K. The unique feature of the temperature dependence of $V_{SC}$ is regardless of microwave powers. Furthermore, the $V_{SC}$ as a function of temperature was also studied by applying different microwave frequencies of $f = 5, 6,$ and 7 GHz at a fixed microwave power of $P_{in} = 25$ mW, as shown in Fig. 4**B**. A similar tendency of temperature dependence of $V_{SC}$ was observed at the different frequencies. In addition, the value of $T_{cross}$ for the sign change of $V_{SC}$ remains the same ($\sim 40$ K) under the different conditions. A reduction of amplitude of $V_{SC}$ is observed with the increase of microwave frequency, which is consistent with the previously reported relationship between $V_{SC}$ and $f$ (*30*). Moreover, $V_{SC}$ was investigated in NiMnSb films with different film thicknesses. In order to make a quantitative comparison, the measured $V_{SC}$ was divided by the corresponding amplitude of FMR absorption, $P_{FMR}$, of the YIG. Figure 4**C** shows the $V_{SC}/P_{FMR}$ as a function of temperature for the NiMnSb films with the thicknesses of $t = 10, 20, 30,$ and 50 nm, as well as for a less-ordered NiMnSb film with the thickness of 20 nm. It is found that the less-ordered sample shows a large $V_{SC}$ in the entire temperature range, which could be attributed to the existence of more scattering centers due to the low structural order in the sample. In addition, the magnitude of $V_{SC}$ decreases with increasing NiMnSb thickness from 10 to 50 nm. Apart from the magnitude of $V_{SC}$, the temperature dependence shows a similar trend for all the samples regardless thickness and ordering structure. Nevertheless, the temperature of $T_{cross}$ for the sign change of $V_{SC}$ reveals a distinct dependence on the film thickness, as shown in Fig. 4**D**. An enlarged view of Fig. 4**C** in the temperature region of below 100 K is also shown in the inset of Fig. 4**D**. The $T_{cross}$ increases from 25 to 75 K with the increase of NiMnSb thickness from 10 to 50 nm. The less-ordered sample shows no sign change of $V_{SC}$ in the measuring temperature range.

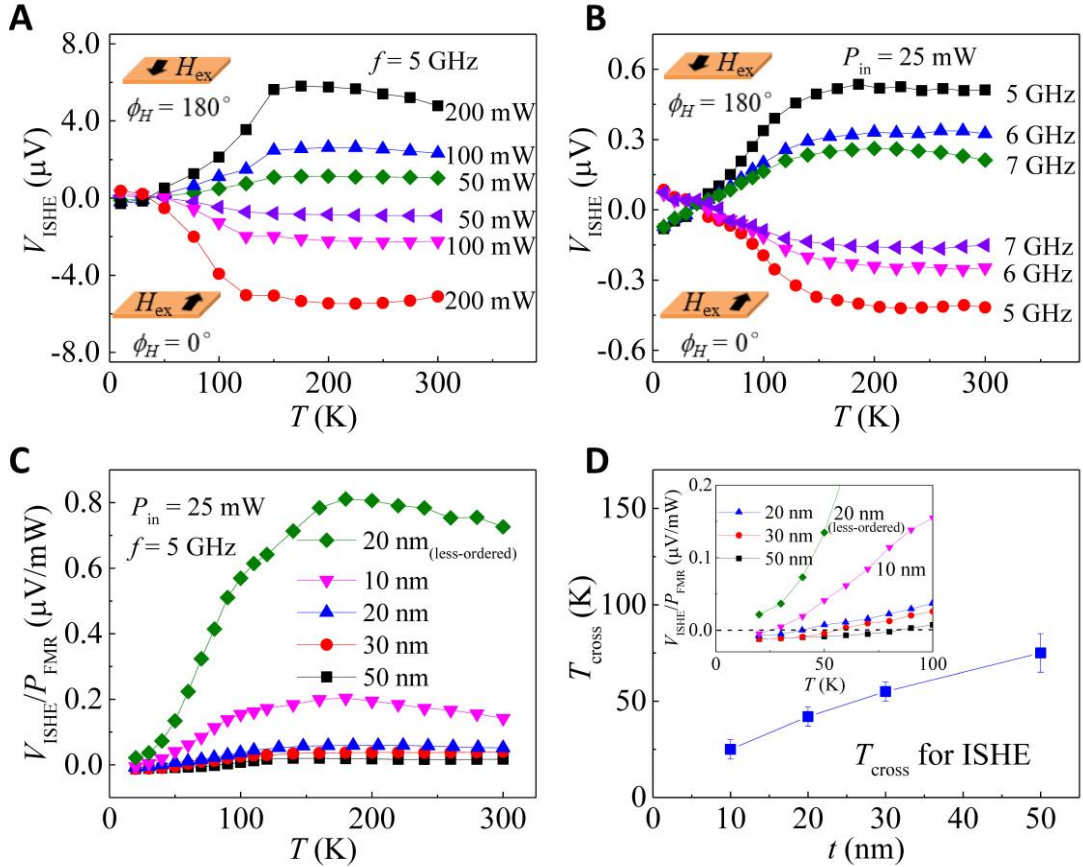

**Fig. 4. Temperature dependence of $V_{SC}$ with varying microwave power, microwave frequency and NiMnSb thickness.** (**A**) $V_{SC}$ as a function of temperature, measured at different microwave powers of $P_{in}$ = 50, 100, and 200 mW with the microwave frequency of $f$ = 5 GHz. (**B**) $V_{SC}$ as a function of temperature at different microwave frequencies of $f$ = 5, 6, and 7 GHz with the microwave power of $P_{in}$ = 25 mW. (**C**) Temperature dependence of $V_{SC}$ divided by the amplitude of FMR absorption of YIG for NiMnSb films with the thickness of 10, 20, 30, and 50 nm. The $V_{SC}$ measured in a less-ordered NiMnSb film with the thickness of 20 nm is also shown for making a comparison. The microwave condition is $P_{in}$ = 25 mW and $f$ = 5 GHz. (**D**) The dependence of cross temperature, $T_{cross}$, for the sign change of $V_{SC}$ on NiMnSb thickness, $t$. Inset is an enlarged view of (**C**) in the temperature region of below 100 K.

**DISCUSSION**

In the following, we theoretically discuss and analyze the observed voltage due to spin-charge conversion, as well as its temperature dependence. Technical details are presented in the Supplemental Material. Firstly, by considering the low temperature regime, the experimental data show a nonvanishing voltage $V_0$ when approaching zero temperature. At $T = 0$ K, it is however impossible to inject a pure spin current − a net spin current in the absence of any charge current − into a perfect HMF: pumping such a current requires establishing an equal-in-magnitude and opposite-in-sign chemical potential imbalance for two different spin species, while a perfect HMF has only one available. Since the pumping source is YIG, a ferrimagnetic insulator (no carriers available for spurious charge injection), this signal must be related to a spin-orbit active interface region, expected to be about 1−2 nm, where the perfect half-metallicity is known to be degraded (*17–22*). Very likely, the interface exhibits a Rashba-like spin-orbit coupling due to structural inversion asymmetry (*31*). The voltage at $T = 0$ K is then attributed to the spin galvanic (or inverse Edelstein) effect (*32–34*), and in the intrinsically dominated case given by

$$V_0 = -\frac{2m\alpha}{enL} j^s, \qquad (2)$$

where $e = |e|$ is the elementary charge, $L$ is the sample length, and $j^s$ is the (three-dimensional) spin current generated by YIG. Furthermore, $\alpha$ is the Rashba coefficient, $m$ the effective electron mass, and $n$ the (two-dimensional) particle density of the interface region, respectively.

Secondly, at finite temperatures $T > 0$ K, even perfectly half-metallic bulk NiMnSb is capable of absorbing a pure spin current. This is because electron-magnon (spin flip) scattering yields a finite Density of States (DOS) for the minority band of NiMnSb (*16, 35–40*),

$$\mathcal{N}^\downarrow(\epsilon_F) \approx \mathcal{N}^\uparrow(\epsilon_F)\left(\frac{T}{T^*}\right)^{3/2}, \qquad (3)$$

where $\mathcal{N}^\uparrow, \mathcal{N}^\downarrow$ are, respectively, the majority (↑) and minority (↓) DOS, while $T^* \sim 10^3$ K is a characteristic temperature corresponding to the magnon energy at the boundary of the Stoner continuum. The resulting absorbed spin current $j^s_{\text{eff}}(T)$ is

$$j^s_{\text{eff}}(T) \approx \frac{\mathcal{N}^\downarrow(\epsilon_F)}{\mathcal{N}^\uparrow(\epsilon_F)} j^s \approx \left(\frac{T}{T^*}\right)^{\frac{3}{2}} j^s. \tag{4}$$

The spin current $j^s_{\text{eff}}(T)$ is then converted to a transverse voltage via the inverse spin Hall effect in bulk NiMnSb due to, for example, extrinsic effects like side-jump and skew scattering with impurities and phonons (*41*), yielding

$$V_{\text{HMF}} = \frac{e\rho\theta_{\text{sH}}\lambda_{sd}L}{2d} \tanh\left(\frac{d}{2\lambda_{sd}}\right) j^s_{\text{eff}}. \tag{5}$$

Here $d$, $\theta_{\text{sH}}$, and $\rho$ are the thickness, spin Hall angle, and (3D) resistivity of the sample, respectively.

Thus, the total voltage due to spin-charge conversion is the sum of the interface and bulk magnon-induced contributions

$$V_{\text{SC}}(T) = V_0 + V_{\text{HMF}}(T). \tag{6}$$

The interfacial spin-charge conversion is taken to be $T$-independent ($V_0$ = const) as a consequence of the following: (i) It is of structural nature and not magnon-limited; (ii) As shown by the experimental data, $|V_0| \ll |V_{\text{HMF}}(T \gtrsim 100~\text{K})|$, so that its possible $T$-dependence is irrelevant for the following analysis. Notice however that the condition $|V_0| \ll |V_{\text{HMF}}(T \gtrsim 100~\text{K})|$ approaches breakdown in the thicker 50-nm sample (see discussion below). The thickness dependent crossing temperature can already be qualitatively explained: Assuming $d \gg \lambda_{sd}$ and $V_{\text{HMF}} \approx$ const. $> 0$ in the high temperature region, an increasing thickness $d$ compresses the total $V_{\text{HMF}}(T) \sim 1/d$ curve, see Eq. (5). For $V_0 < 0$ (see inset of Fig. 4D), the crossing temperature thus increases with increasing thickness as sketched in Fig. 5.

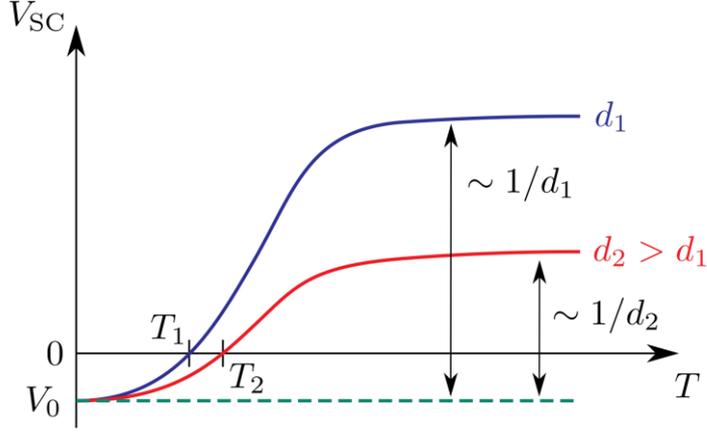

**Fig. 5. Sketch of the voltage due to spin-charge conversion as a function of temperature for two thicknesses:** $d_1$ (blue) and $d_2$ (red) ($d_2 > d_1$), with the crossing temperatures $T_1$ and $T_2$ ($T_2 > T_1$), respectively. The green dashed line is the interfacial contribution $V_0$.

In addition to the temperature-dependent effective pumping spin current, Eq. (4), the resistivity $\rho$, spin Hall angle $\theta_{sH}$, and spin diffusion length $\lambda_{sd}$, all appearing in Eq. (5), are temperature dependent as well. Let us start from the latter,

$$\lambda_{sd} = \sqrt{\frac{2D_\uparrow D_\downarrow \tau_s}{D_\uparrow + D_\downarrow}}. \qquad (7)$$

The spin diffusion length depends on the diffusion constants of majority and minority spin electrons, $D_\uparrow$ and $D_\downarrow$, respectively, and on the spin-flip relaxation time $\tau_s$. Assuming $D_\uparrow \gg D_\downarrow$, and that $D_\downarrow$ and $\tau_s$ are proportional to $1/\mathcal{N}^\downarrow$, the spin diffusion length is estimated as

$$\lambda_{sd} \approx \sqrt{2D_\downarrow \tau_s} \sim \left(\frac{T^*}{T}\right)^{\frac{3}{2}}. \qquad (8)$$

From Eq. (4), note that the product $\lambda_{sd} j_{\text{eff}}^s$ in Eq. (5) is independent of temperature.

The spin Hall angle can be separated into majority and minority band contributions, $\theta_{sH}^\uparrow$ and $\theta_{sH}^\downarrow$, respectively, and the same is done for the corresponding spin Hall conductivities, $\sigma_{sH}^\uparrow$ and $\sigma_{sH}^\downarrow$. According to *Ref.* 16, $\sigma_{sH}^\downarrow$ is negligible compared to $\sigma_{sH}^\uparrow$, so that $\theta_{sH} \approx \theta_{sH}^\uparrow$. The $T$-dependence of the spin Hall angle of a metallic system was discussed by Karnad *et al.* (*42*), where intrinsic spin-orbit coupling and skew scattering dominate. We follow this treatment, introducing, however, in addition to skew scattering, a phenomenological contribution which is lifetime and thus temperature independent. Its microscopic origin could be extrinsic, e.g., due to side-jump (*43*), or intrinsic, e.g., arising from Fermi sea contributions of the kind relevant for the anomalous Hall effect (*44*, *45*). This phenomenological contribution to the corresponding spin Hall conductivity is denoted by $\sigma_{sH}^*$, while the skew scattering contribution at temperature $T$ is denoted, as usual, by $\sigma_{sH}^{ss,T}$. Finally, the resistivity $\rho(T)$ of our samples is known and appears in Fig. 6**A**. The overall voltage is thus described by

$$V_{SC}(T) = V_0 + \frac{AP_{FMR}}{d}(1 + BT)\tanh(dCT^{3/2}), \tag{9}$$

which serves as a fitting equation with the parameters $A$, $B$, and $C$ to the experimental data, Fig. 6**B**, where $V_0/P_{FMR} = -0.02$ µV/mW (See details in the Supplemental Material). Notice that this is not a 3-parameter fit, since $A$, $B$, and $C$ are not independent. $B$ is further constrained by the measured resistivity values via the known parameter $\gamma$, see Fig. 6**A** and details in the Supplemental Material. The parameter $B$ allows us to extract the ratio between the scattering-independent and the skew scattering contributions at zero temperature, see Table I. From $C$, we obtain the spin-diffusion lengths for each layer thickness. As shown in Table I, $\lambda_{sd}$ at room temperature has reasonable values in the range of nanometers.

TABLE I. The spin-diffusion length at room temperature (300 K), and the ratio $\sigma_{sH}^*/\sigma_{sH}^{ss,0}$ for the various thicknesses of the NiMnSb films, respectively. These are obtained by the fit of Eq. (9) to the experimental data as shown in Fig. 6**B** for $V_0/P_{FMR} = -0.03$ µV/mW, $-0.02$ µV/mW, $-0.01$ µV/mW from left to right. Note that $\sigma_{sH}^*$ is temperature independent by definition, and that $\sigma_{sH}^{ss,0}$ denotes the zero-temperature value of the skew scattering contribution.

| $d$ | $\lambda_{sd}$ (300 K) [nm] | $\sigma_{sH}^*/\sigma_{sH}^{ss,0}$ |
|---|---|---|

| | | | | | | |
|---|---|---|---|---|---|---|
| 50 nm | 9.47 | 14.03 | 19.65 | −0.16 | −0.27 | −0.29 |
| 30 nm | 4.74 | 6.32 | 8.31 | 0.09 | −0.05 | −0.12 |
| 20 nm | 4.21 | 5.48 | 7.27 | −0.23 | −0.32 | −0.39 |
| 10 nm | 2.82 | 3.07 | 3.37 | −0.60 | −0.62 | −0.63 |

Equation (9) fits very well the experimental data of the 10, 20, 30, and 50 nm thick samples. The slightly reduced accuracy in the case of 50 nm data can be explained by the decrease of the bulk contribution, $V_{HMF} \sim 1/d$ according to Eq. (5). Particularly, $|V_{HMF}| \approx |V_0|$ even at high temperature for the 50 nm sample, so that the $T$-dependence of the interfacial contribution $V_0$ could play a considerable role. The overall good match between data and fitted curves supports our argument that the total signal is the sum of: (i) a mostly temperature independent interfacial contribution $V_0$; (ii) a thickness-dependent bulk one mainly limited by minority states due to thermally excited magnons, with spin diffusion length $\sim T^{-3/2}$.

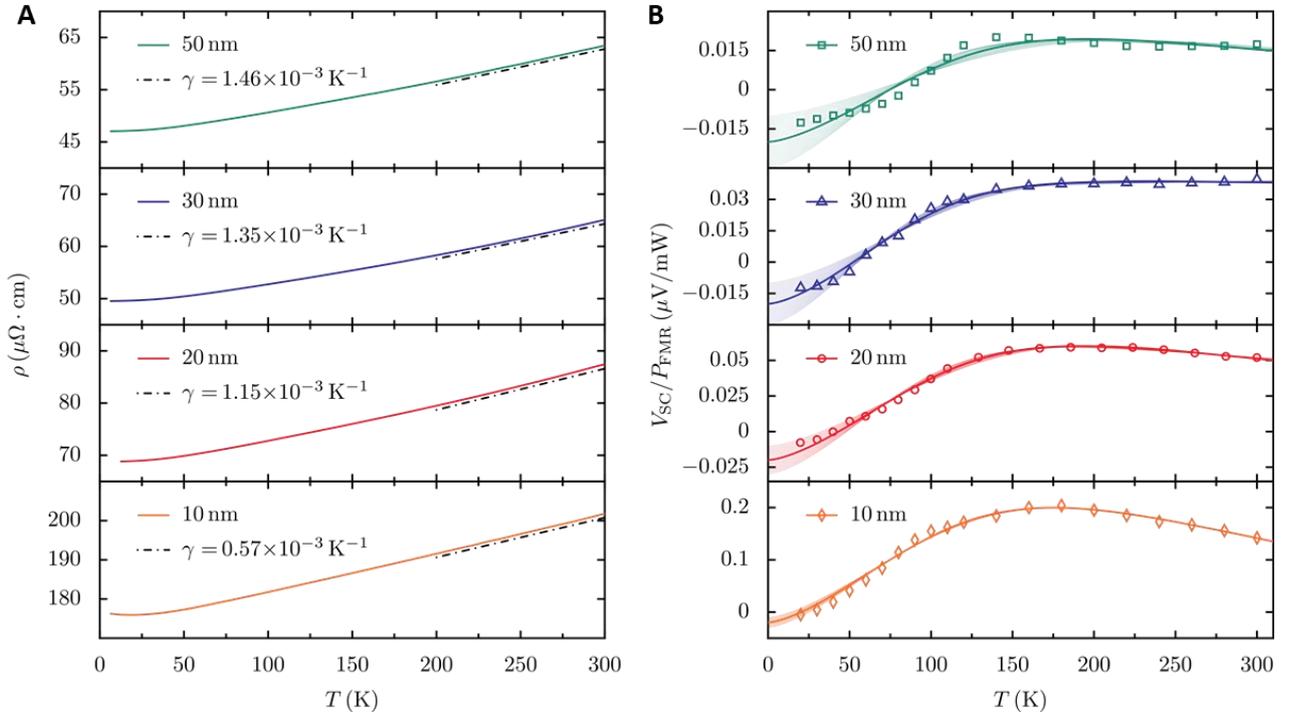

**Fig. 6. Resistivity of NiMnSb films and the voltage due to spin-charge conversion as a function of temperature.** (A) The dependence of the resistivity on temperature for the NiMnSb films with $d =$ 10 nm, 20 nm, 30 nm, 50 nm. The linear $T$-dependence ($\sim \gamma T$) in the high temperature regime is

indicated by a dash dotted line with the respective value of γ for each thickness. (**B**) The voltage signal divided by the amplitude of FMR absorption of YIG against the temperature for the various thicknesses of the NiMnSb layer. The symbols represent the experimental data and the solid line shows their fitted curve with use of Eq. (9) with $V_0/P_{FMR}$ = −0.02 µV/mW, respectively. The shaded areas show the range of fitting curves for −0.03 µV/mW < $V_0/P_{FMR}$ < −0.01 µV/mW.

In conclusion, the spin-charge conversion in NiMnSb alloy films was investigated through the injection of spin current from YIG by spin pumping. It was observed that the voltage due to spin-charge conversion showed an unusual temperature dependence and a sign change at low temperature depending on film thickness and ordering structure. The temperature dependent behavior of the voltage in the NiMnSb films was further analyzed by the spin-charge conversion contributed by interface and bulk effects. This study may contribute a way for efficient spin-charge conversion in FMs for spintronic devices.

**MATERIALS AND METHODS**

The NiMnSb alloy films were deposited on YIG substrates by a co-sputtering method from Ni and MnSb targets in an ultrahigh vacuum magnetron sputtering system with a base pressure of ~ 1 × 10$^{-7}$ Pa. 1.5-nm-thick AlO$_x$ was subsequently deposited on the NiMnSb films for protection. The stoichiometric composition of the NiMnSb films was confirmed to be Ni$_{1.01}$Mn$_{0.98}$Sb$_{1.01}$ by inductively coupled plasma (ICP) analysis. The samples with high ordering structure were deposited at a substrate temperature of 500 °C while the sample with low order was grown at room temperature. The structural properties of NiMnSb films were characterized by x-ray diffraction (XRD) with Cu $K_\alpha$ radiation ($\lambda$ = 0.15418 nm). AMR effect was measured at 10 and 300 K by a physical properties measurement system (PPMS). The YIG-substrate//NiMnSb device was placed on a coplanar waveguide where the microwave was applied. Two terminal electrodes were attached on the device for measuring the voltage. The magnetic field was applied in the plane of the films and perpendicular to the direction across the two electrodes. The power of microwave was varied from 25 to 200 mW and the frequency was changed from 5 to 7 GHz. The voltage signal was detected by using lock-in techniques and the voltage measurement was performed in PPMS at low and room temperatures.

**SUPPLEMENTAL MATERIAL**

Experimental results on the FMR spectrum of NiMnSb and the voltage signals due to the NiMnSb resonance:

Fig. S1. Typical FMR spectra and voltage signals in the YIG//NiMnSb (20 nm) sample.

Technical details of the theory of the spin-charge conversion:

A. Interfacial contribution to the voltage

B. Inverse spin Hall voltage of the HMF

C. Temperature dependence of the spin Hall angle

D. Fit to the experimental data

**Acknowledgments:** We thank S. Maekawa, Y. Ohnuma, J. Xiao, V. L. Grigoryan, L. Chioncel, and K. Samwer for fruitful discussions. **Funding:** This work was partially supported by the KAKENHI (S) (No. 18H05246) and the Grant-in-Aid for Young Scientists B (No. 17K14652) from the Japan Society for the Promotion of Science (JSPS), and the Advanced Storage Research Consortium (ASRC). ZQ acknowledges the support from the 'Fundamental Research Funds for the Central Universities (DUT17RC(3)073). ST, CG, and UE acknowledge the support from the German Science Foundation (DFG) through TRR 80 and SFB 1277. **Author contributions:** KT supervised this study. ZW and ZQ fabricated all the films, measured the samples. ZW, ZQ, TS, DH, TK, KT, and ES analyzed the data. ST, CG, and UE developed the theoretical discussion. All authors discussed the results and prepared the manuscript. **Competing interests:** The authors declare that they have no competing interests. **Data and materials availability:** All data needed to evaluate the conclusions in the paper are present in the paper and/or the Supplemental Material. Additional data related to this paper may be requested from the authors.

# SUPPLEMENTAL MATERIAL

**Part 1. Experimental results on the FMR spectrum of NiMnSb and the voltage signals due to the NiMnSb resonance**

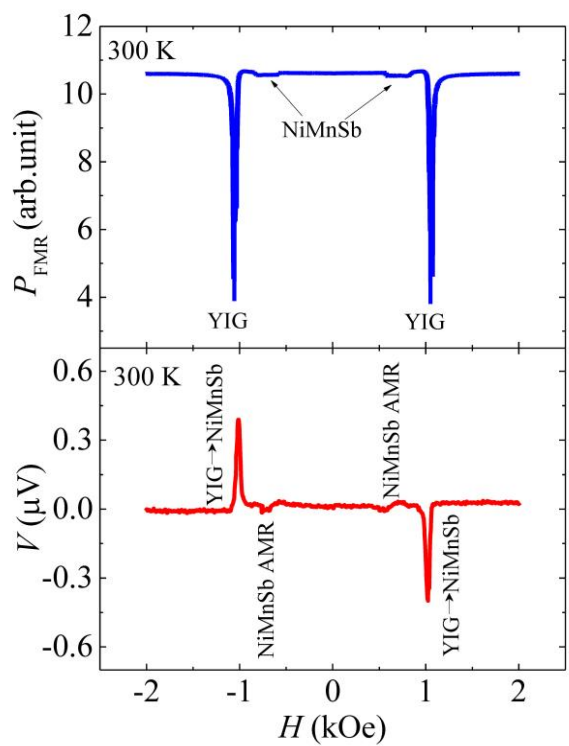

**Fig. S1. Typical FMR spectra and voltage signals in the YIG//NiMnSb(20 nm) sample.** The applied microwave power and frequency are 25 mW and 5 GHz, respectively.

The FMR absorption spectrum of NiMnSb layer and the electric voltage at the NiMnSb resonance were also observed, as shown in Fig. **S1**. The large FMR absorption shown at $H = \pm 1100$ Oe is corresponding to the FMR resonance of YIG (*46*, *47*) while the very small magnitude of FMR absorption shown between the FMR absorptions of YIG is attributed to the NiMnSb layer. In addition, the voltage signals exhibit the consistence with the FMR resonances. The voltage peaks at the NiMnSb resonance is mainly due to the FMR-induced AMR in NiMnSb, and the similar behavior was also observed in NiFe system (*11*, *28*). The separation of resonance field positions between the YIG and NiMnSb indicates the distinction of the electric voltage signals due to spin pumping from the YIG and FMR in NiMnSb.

**Part 2. Technical details of the theory of the voltage due to spin-charge conversion**

**A. Interfacial contribution to the voltage**

We explain the interfacial contribution to the voltage, $V_0$, by a Rashba-like spin-orbit coupling due to the structural inversion asymmetry at the YIG/NiMnSb interface. The Rashba Hamiltonian reads

$$H_R = -\frac{\alpha}{\hbar}(\mathbf{p} \times \boldsymbol{\sigma}) \cdot \hat{\mathbf{z}} \qquad (S1)$$

where $\alpha$ is the Rashba parameter, $\mathbf{p}$ is the momentum of the interfacial electrons, $\boldsymbol{\sigma} = (\sigma^x, \sigma^y, \sigma^z)$ is the vector of Pauli matrices, and $\hat{\mathbf{z}}$ is the unit vector perpendicular to the interface. A nonequilibrium spin accumulation, induced in our case by means of an injected spin current from the YIG, is converted by spin-orbit coupling (*34*) into a charge current within the interface region. This is the spin galvanic (or inverse Edelstein) effect (*32–34*). According to *Ref.* 32, see Eq. (21) therein, in the intrinsically dominated case, i.e., when extrinsic spin-orbit coupling can be neglected, the charge current is given by

$$j_{2D}^c = -\frac{e\alpha\tau_{2D}}{\hbar}j^s, \qquad (S2)$$

where $e = |e|$ is the elementary charge and $\tau_{2D}$ is the momentum relaxation time within the interface region. Note that $j^s$ has the dimensionality of a three-dimensional particle current since it refers to the spin current, generated by the YIG, which flows perpendicular to the interface, whereas $j_{2D}^c$ is a two-dimensional charge current which flows in-plane in the interface region. In an open-circuit situation we have $j_{2D}^c = \sigma_{2D}LV_0$, where $L$ is the length of the sample in current direction, and $\sigma_{2D} = e^2 n\tau_{2D}/m$ is the Drude conductivity of the interface with effective electron mass $m$ and particle density $n$. The corresponding voltage is therefore given by

$$V_0 = -\frac{2m\alpha}{enL}j^s. \qquad (S3)$$

Extrinsic spin-orbit coupling can be included (*33*), leading to additional and possibly temperature dependent contributions to $V_0$. As stated in the main text, the latter appear negligible in our context and will not be considered.

## B. Inverse spin Hall voltage of the HMF

We derive Eq. (5) in the main manuscript, i.e., the inverse spin Hall voltage of the half-metallic ferromagnet (HMF) due to a spin current $j^s_{\text{eff}}$ which is pumped into the (bulk) NiMnSb layer. We rely on the standard drift-diffusion approach, specialized to our system. The geometry is assumed as follows: the YIG/NiMnSb interface lies in the $xy$ plane at $z = 0$ with the NiMnSb layer stretching from $z = 0$ to $z = d$ in height, and from $x = 0$ to $x = L$ in length. In order to describe the spin transport in the HMF, we start from the 'continuity' equation,

$$\partial_t s + \nabla_z j^s_z = -\frac{1}{\tau_s} s, \qquad (S4)$$

for the nonequilibrium spin density $s = n^\uparrow - n^\downarrow$ with the spin current in $z$ direction $j^s_z = j^\uparrow_z - j^\downarrow_z$, and the spin-flip relaxation rate $1/\tau_s$. Here, $n^\uparrow(n^\downarrow)$ is the majority-spin (minority-spin) particle density (per volume) induced by the spin pumping. The majority- and minority-spin particle currents $j^\uparrow_z$ and $j^\downarrow_z$ are given as diffusive currents,

$$j^{\uparrow,\downarrow}_z = -D_{\uparrow,\downarrow} \nabla_z n^{\uparrow,\downarrow}, \qquad (S5)$$

where $D_{\uparrow,\downarrow}$ is the diffusion constant per spin channel. Inserting Eq. (S5) into Eq. (S4) we easily find the general solution for $n^{\uparrow,\downarrow}$ and thus also for $j^{\uparrow,\downarrow}_z$, using again Eq. (S5). For the spin current, we take the boundary conditions as

$$j^s_z(z = 0) = j^s_{\text{eff}}, \quad j^s_z(z = d) = 0, \qquad (S6)$$

where the first one corresponds to the spin pumping scenario and the second one to spin conserving scattering at the surface. In addition, no charge flows along $z$, i.e., we have open-circuit conditions $j_z^c = -e(j_z^\uparrow + j_z^\downarrow) = 0$. This yields the mean injected particle current per spin channel

$$\bar{j}_z^\uparrow = -\bar{j}_z^\downarrow = \frac{1}{d}\int_0^d dz\, j_z^\uparrow = \frac{\lambda_{sd}}{2d}\tanh\left(\frac{d}{2\lambda_{sd}}\right) j_{\text{eff}}^s, \quad (S7)$$

where $\lambda_{sd} = \sqrt{2D_\uparrow D_\downarrow \tau_s/(D_\uparrow + D_\downarrow)}$ is the diffusion length in terms of the diffusion constants and the spin-flip relaxation time.

Due to the inverse spin Hall effect, the injected spin current induces a particle current (*48*),

$$j_x^\uparrow = \sigma^\uparrow E_x - \theta_{\text{SH}}^\uparrow j_z^\uparrow, \qquad j_x^\downarrow = \sigma^\downarrow E_x + \theta_{\text{SH}}^\downarrow j_z^\downarrow, \quad (S8)$$

where $\sigma^{\uparrow,\downarrow}$ are the majority and minority longitudinal (particle) conductivities, and $\theta_{\text{SH}}^{\uparrow,\downarrow}$ is the spin Hall angle per spin channel. In an open-circuit situation, i.e., $j_x^c = -e(j_x^\uparrow + j_x^\downarrow) = 0$, the electric field is given by

$$E_x = e\rho \left(\theta_{\text{SH}}^\uparrow j_z^\uparrow - \theta_{\text{SH}}^\downarrow j_z^\downarrow\right), \quad (S9)$$

where $\rho = -e/(\sigma^\uparrow + \sigma^\downarrow)$ is the resistivity. The inverse spin Hall voltage of the HMF is then obtained by multiplying Eq. (S9) with the length $L$, performing the thickness average, and inserting Eq. (S7),

$$V_{\text{HMF}} = \frac{L}{d}\int_0^d dz\, E_x = \frac{e\rho\theta_{\text{SH}}\lambda_{sd}L}{2d}\tanh\left(\frac{d}{2\lambda_{sd}}\right) j_{\text{eff}}^s, \quad (S10)$$

where $\theta_{\text{SH}} = \theta_{\text{SH}}^\uparrow + \theta_{\text{SH}}^\downarrow$ is the total spin Hall angle.

### C. Temperature dependence of the spin Hall angle

Next, we discuss the temperature dependence of the spin Hall angle due to side-jump and skew scattering. Since $\sigma_{\text{SH}}^\downarrow \ll \sigma_{\text{SH}}^\uparrow$, see *Ref.* 16, one has $\theta_{\text{SH}} \approx \theta_{\text{SH}}^\uparrow$, and we will therefore omit the superscript ↑ in the following.

According to the *Ref.* 41 (Eq. (7) therein), the temperature dependent skew-scattering spin Hall

conductivity reads

$$\sigma_{\text{sH}}^{\text{ss},T} = \frac{\sigma_{\text{sH}}^{\text{ss},0}}{[1+\eta(T)]^2}[1+\eta_2(T)], \qquad (S11)$$

where $\sigma_{\text{sH}}^{\text{ss},0}$ is the skew-scattering spin Hall conductivity at zero temperature, $\eta(T)$ is defined by the temperature dependence of $\rho = \rho_0(1+\eta(T))$, and $\eta_2(T)$ is a finite temperature contribution due to phonon skew scattering (49). Note that $\eta_2(T) \sim (T/T_F)\eta(T) \ll \eta(T)$ for temperatures well below the Fermi temperature. The total spin Hall conductivity is then obtained by adding the temperature independent side-jump contribution $\sigma_{\text{sH}}^{\text{sj}}$,

$$\sigma_{\text{sH}}(T) = \sigma_{\text{sH}}^{\text{sj}} + \frac{\sigma_{\text{sH}}^{\text{ss},0}}{[1+\eta(T)]^2}[1+\eta_2(T)]. \qquad (S12)$$

The spin Hall angle is now defined by the ratio $\sigma_{\text{sH}}/(\sigma^\uparrow + \sigma^\downarrow)$, and with use of $\sigma^\uparrow + \sigma^\downarrow = -e/\rho$, see Eq. (S9), we find

$$\theta_{\text{sH}}(T) = -\frac{1}{e}\rho(T)\sigma_{\text{sH}}(T). \qquad (S13)$$

We insert Eq. (S12) into Eq. (S13), and by employing $\rho = \rho_0(1+\eta(T))$ we obtain

$$\theta_{\text{sH}}(T) = \frac{\rho_{\text{sH}}^0}{\rho(T)}[1+\tilde{\eta}(T)], \qquad (S14)$$

with

$$\rho_{\text{sH}}^0 = -\frac{1}{e}\rho_0^2\left(\sigma_{\text{sH}}^{\text{ss},0} + \sigma_{\text{sH}}^{\text{sj}}\right),$$

$$\tilde{\eta}(T) = \frac{2\eta(T)}{1+\sigma_{\text{sH}}^{\text{ss},0}/\sigma_{\text{sH}}^{\text{sj}}} + \frac{\eta^2(T)}{1+\sigma_{\text{sH}}^{\text{ss},0}/\sigma_{\text{sH}}^{\text{sj}}} + \frac{\eta_2(T)}{1+\sigma_{\text{sH}}^{\text{sj}}/\sigma_{\text{sH}}^{\text{ss},0}}. \qquad (S15)$$

The experiments are performed in the temperature range $0 < T \leq 300$ K in highly-disordered samples. Thus, we are away from the "ultra-clean" regime of *Ref.* 41 and well below the Fermi temperature, and we will neglect $\eta_2(T) \ll \eta(T)$. Indeed, at the moment one can only speculate about the increasing importance of phonons, and in particular of phonon skew scattering, with increasing temperature. On the one hand, the importance is suggested by the low Debye temperature of NiMnSb, especially of its surface modes affecting the interfacial spin galvanic conversion (*50*). On the other hand, the magnon-limited bulk signal appears dominant in all samples considered in our study. Notice also that $\eta(T) \ll 1$ for the experimental data (Fig. 6A in the main manuscript), so that we can approximate

$$\tilde{\eta}(T) \approx \frac{2\eta(T)}{1 + \sigma_{\text{sH}}^{\text{ss,0}}/\sigma_{\text{sH}}^{\text{sj}}}. \qquad (S16)$$

As discussed in the main text in connection with Eq. (9), there may be additional intrinsic contributions which we take into account in the following by replacing $\sigma_{\text{sH}}^{\text{sj}}$ by $\sigma_{\text{sH}}^{*}$. Note that $\sigma_{\text{sH}}^{*}$ is $T$ independent.

### D. Fit to the experimental data

In order to connect the total voltage,

$$V_{\text{SC}}(T) = V_0 + V_{\text{HMF}}(T), \qquad (S17)$$

with the fit equation (9) in the main text, let us recall

$$\lambda_{sd} j_{\text{eff}}^s = \text{const},$$

$$\lambda_{sd} \approx \sqrt{2D_\downarrow \tau_s} = \lambda_{sd}^* \left(\frac{T^*}{T}\right)^{\frac{3}{2}}, \qquad (S18)$$

where is $\lambda_{sd}^*$ is the spin diffusion length at $T = T^*$. We insert Eq. (S14) into Eq. (S10), and with the estimations (S16) and (S18), the total voltage, Eq. (S17), becomes

$$V_{\text{SC}}(T) = V_0 + \frac{e\rho_{\text{sH}}^0 L \lambda_{sd} j_{\text{eff}}^s}{2d} \left[1 + \frac{2\eta(T)}{1 + \sigma_{\text{sH}}^{\text{ss,0}}/\sigma_{\text{sH}}^{*}}\right] \tanh\left(\frac{d}{2\lambda_{sd}^*} \left(\frac{T}{T^*}\right)^{3/2}\right). \qquad (S19)$$

As shown in Fig. 6A in the main text, the resistivity is linear in the temperature already well below

100 K, $\eta(T) \approx \gamma T$. We determine $\gamma$ from the measured resistivity curves, and identify the fitting parameters appearing in Eq. (9) in the main text with

$$A = \frac{e\rho_{sH}^0 L \lambda_{sd} j_{eff}^s}{2P_{FMR}}, \qquad (S20)$$

$$B = \frac{2\gamma}{1 + \sigma_{sH}^{ss,0}/\sigma_{sH}^*}, \qquad (S21)$$

$$C = \frac{1}{2\lambda_{sd}^* T^{*3/2}}. \qquad (S22)$$

Hence, the ratio of the scattering-independent to skew scattering spin Hall conductivities and the temperature-dependent spin diffusion length can be extracted with the fitting parameters as

$$\frac{\sigma_{sH}^*}{\sigma_{sH}^{ss,0}} = \frac{B}{2\gamma - B}, \qquad (S23)$$

$$\lambda_{sd}(T) = \frac{T^{-3/2}}{2C}, \qquad (S24)$$

respectively.